\documentclass[prl,twoside,showpacs,twocolumn,floatfix]{revtex4-1}

\usepackage{amssymb}
\usepackage{amsfonts}
\usepackage{amsmath}
\usepackage[dvips]{graphicx}
\usepackage{enumerate}
\usepackage{latexsym}
\usepackage{soul}
\usepackage[dvips,usenames,dvipsnames]{color}
\newcommand{\Deltau}{\Delta}
\newcommand{\Deltae}{\partial e}

\begin{document}
\sloppy
\title{Fluctuation theorems for an asymmetric rotor in a granular gas}
\author{S. Joubaud$^{1,2}$, D. Lohse$^{1}$ and D. van der Meer$^{1}$}
\affiliation{$^{1}$Physics of Fluids group, University of Twente, P.O. Box 217, 7500 AE Enschede, The Netherlands\\
$^{2}$Laboratoire de Physique de l'\'Ecole Normale Sup\'erieure de Lyon, CNRS \& Universit\'e de Lyon, F-69364 Lyon, France}
\date{\today}

\begin{abstract}
We investigate the validity of fluctuation theorems for an asymmetric rotor experiment in a granular gas. A first state, with a Gaussian distribution of the angular velocity, is found to be well described by a first order Langevin equation. We show that fluctuation theorems are valid for the injected work and for the total entropy production. In a second state the angular velocity distribution is double-peaked due to a spontaneous symmetry breaking: A convection roll develops in the granular gas, which strongly couples to the rotor. Surprisingly, in this case similar symmetry relations hold, which lead to a good prediction for the height ratio of the two peaks.
\end{abstract}


\maketitle
Under rather general conditions, the thermodynamic description of non-equilibrium systems is constrained by the consequences of the fluctuation relations~\cite{Kurchan2007}. 
In particular, the fluctuation theorem (FT) provides, for a system in a non-equilibrium steady state, a quantitative symmetry relation between the probability of having a positive fluctuation for the entropy production in a time $\tau$ and a corresponding negative one:
\vspace{-0.2cm}
\begin{equation}
\ln\left(\frac{P(\sigma_{\tau}=+a)}{P(\sigma_{\tau}=-a)}\right) = \beta a\quad \forall\, a\quad \textrm{for} \,\tau\gg\tau_c
\label{FT:eq}
\end{equation}
\vspace{-0.5cm}

\noindent where $\tau_c$ represents the largest characteristic time of the system and $\beta$ a prefactor~\cite{Evansetal93,Gallavotti1995, Evans-Searles, footnote0}. In general, the hypotheses used to prove this theorem are not verified experimentally and therefore, it is not clear whether Eq.~(\ref{FT:eq}) holds or not. Experimental tests of FTs have been mostly performed on stochastic systems in contact with a thermal bath. In such experiments, a FT is valid for the injected work into the system or the total entropy production and the coefficient $\beta$ is directly related to the temperature of the thermal bath $\beta=1/k_BT$~\cite{Ciliberto2010}. 

Experimental tests are particularly scarce and inconclusive for another type of systems for which FTs are believed to hold, namely non-thermal dynamical systems. In these systems, the strong fluctuations come from the non-linear interaction of many degrees of freedom of a dissipative system. An interesting example of such a system is a granular gas, which by its nature is already out-of-equilibrium due to the dissipative character of the inelastic collisions. Experiments searching for FTs in granular systems have been performed~\cite{Menon04}, but their interpretation remains unclear~\cite{Ciliberto2010}. It is a challenge to study FTs in such a system and to test to what extent the fluctuation relations can be satisfied.

In this letter we will study FTs for a rotor immersed in a granular gas. Eshuis {\em et~al.}~\cite{Eshuis10} showed that such a system not only exhibits Brownian-like dynamics comparable to a thermal system, but also a state in which symmetry is spontaneously broken, and for which there exists no thermal analogue. Here, we will show that FTs are relevant to {\em both} states.      

\emph{Experiment} -- The rotor is composed of four vanes ($25 \, \times \, 60$~mm$^2$ each, made from a single piece of stainless steel), precisely balanced around an axis which is connected to the container by a low-friction ball bearing. The granular heat bath consists of glass beads of diameter $d=4$~mm (density $\rho = 2600$~kg/m$^3$), which are fluidized by vertical vibrations of the bottom ($z=0$ at rest) with amplitude $a$ and frequency $f$ such that the grains interact dissipatively with the vanes. We present the results obtained for a system of $N = 1000$ particles ($500$ and $2000$ particles have also been used) and for two different heights of the axis above the bottom ($h=51$~mm and $h=75$~mm). The natural dimensionless control parameter of the granular heat bath is the shaking strength $S=4\pi^2f^2a^2/(gh)$, which represents the ratio of the typical kinetic energy injected into the system by the vibrating bottom and the potential energy of the particles at the height of the axis of the rotor. The granular temperature $T_{\rm g}$ is defined as the mean of the kinetic energy fluctuations per particle. The angular position $\theta$ of the vanes is measured using an optical angle encoder, with an accuracy of $1.9\cdot10^{-7}$~rad, at a sampling rate of $1000$~Hz (larger than the typical collision rate, which is about $100$~Hz). After the system has reached a steady state, we start to record the position for typically $15$~min, which is about $10^3$ times the relaxation time of the system. We repeat experiments several times with the same conditions ($S$ and $h$). An asymmetry is voluntarily introduced in the system by coating the left hand side of each vane with rubber tape. This side will therefore be softer, diminishing its coefficient of normal restitution. As a results, vanes are expected to preferentially rotate in  counter-clockwise direction (positive $\theta$). Such a system behaves like a ratchet, as explained in~\cite{Talbot10,vdb07,Puglisi07,Eshuis10,Cleuren2008,Balzan2011}. 

\begin{figure}
\centering
\includegraphics[width=0.9\linewidth]{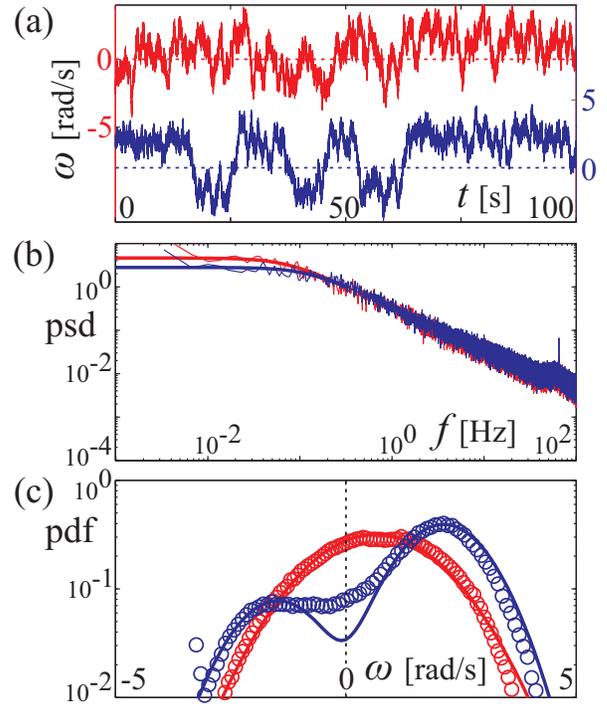}
\caption{\small (color online) (a) Typical time series of the angular velocity $\omega$ for state I [$S=2.78$ and $h=75$~mm (top, red)] and for state II [$S=2.15$ and $h = 51$~mm (bottom, blue)]. (b) Corresponding power spectral densities of the fluctuations $\delta\omega$, which for both cases virtually overlap and are fitted well by a Lorentzian. (c) Corresponding probability distribution functions of $\omega$. For state I (red symbols) the pdf is well fitted by a Gaussian (red line). For state II (blue symbols) there is a pronounced double peak; the blue line is the result of Eqs.~(\ref{double_gauss}) and~(\ref{eq_ratio}).} 
\label{velocity_pdf-psd}
\end{figure}

The dynamics of the system strongly depends on the experimental parameters (height of the vanes $h$, shaking strength $S$, and number of particles). Fig.~\ref{velocity_pdf-psd}a presents two time series of the angular velocity $\omega=\dot{\theta}$ of the rotor: State I has an angular velocity which fluctuates around a non-zero positive mean value (due to the symmetry-breaking coating). In state II, the angular velocity is either positive or negative and there are switches from one state to the other one. This is a state in which, in addition to the symmetry-breaking coating, there is a spontaneously broken symmetry, which can be explained from the development of a convection roll in the granular gas stabilizing the motion of the vanes~\cite{Eshuis10}. These two states will now be studied in detail.

\emph{State I} -- In the first state, the dynamics of the angular velocity $\omega$ can be decomposed into 
a constant positive average value $\langle \omega \rangle$ and strong fluctuations $\delta \omega\equiv \omega-\langle \omega \rangle$ around this mean. To shed more light on the dynamics and statistics, the power spectral density (psd) of $\delta\omega$ and the probability density function (pdf) of $\omega$ are plotted in Figs.~\ref{velocity_pdf-psd}b and c. The Lorentzian shape of their psd and their Gaussian distribution indicate that the fluctuations $\delta \omega$ are identical to those in the absence of the ratchet effect, as expected for a linear situation. The dynamics is therefore likely to be well described by a first order Langevin equation
\begin{equation}
I\frac{{\rm d}\omega}{{\rm d}t}=-\gamma \omega + M_{\Deltae} + \eta\,,
\label{Langevin_model}
\end{equation}
where $I$ ($=7.2\cdot10^{-5}$~kg$\,$m$^2$) is the angular moment of inertia, $\gamma$ a viscous drag coefficient, and $\eta$ a stochastic noise, $\delta$-correlated in time, due to the random collisions between the particles and the rotor. This description is compatible with kinetic theory in the limit in which the mass of the ratchet is much larger than the mass of the particles~\cite{Cleuren2008}. 
The measurement of the variance of the fluctuations, $T_{\rm r}/I$, and the cut-off frequency of the psd, $f_{\rm c} = \gamma/(2\pi I)$ (related to the relaxation time of the system, $\tau_0=1/(2\pi f_c)$) provides the value of the viscosity $\gamma\approx 7.0\cdot10^{-5}$ ~kg$\,$m$^2$/s and the ratchet temperature $T_{\rm r} \approx 0.115$~mJ, which is close to the granular temperature~\cite{footnote1}. 
The ratchet effect is described by adding a constant torque $M_{\Deltae}$ to the Langevin equation~\cite{Cleuren2008}, which is identified with $\gamma\langle \omega\rangle$ ($\approx 4.5\cdot10^{-5}$ Nm) by time averaging of Eq.~(\ref{Langevin_model}). The subscript $\Deltae$ points to the fact that $|M_{\Deltae}|$ is an increasing function of $|\Deltae|$, the difference in normal restitution coefficient of the coated and uncoated side of each vane. For the symmetric system, i.e., without coating, $\langle\omega\rangle = 0$ and thus $M_{\Deltae}=0$. 
  
\begin{figure*}
 \begin{center}
\includegraphics[width=1\linewidth]{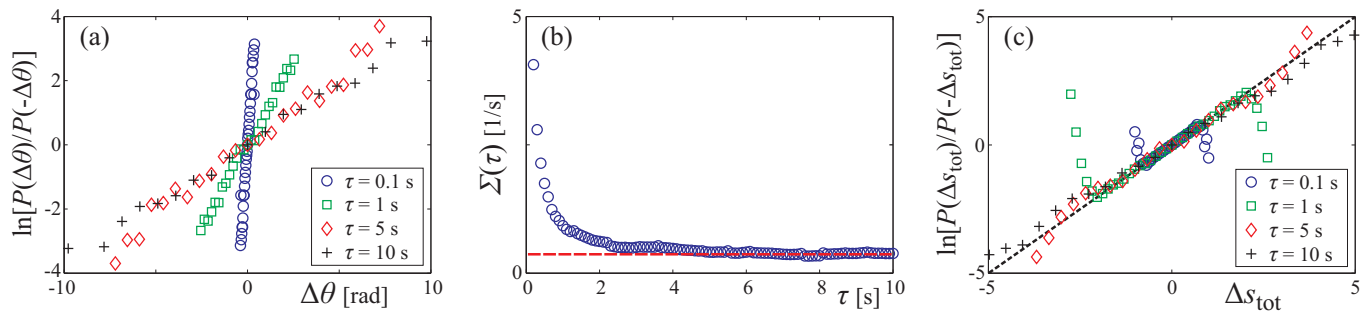}
\caption{\small (color online) State I [$S = 2.78$ and $h = 75$~mm]: (a) The symmetry function $\ln(P(\Deltau \theta)/P(-\Deltau \theta))$ versus $\Deltau \theta$ for different values of the time interval $\tau$. All are linear with $\Deltau \theta$, such that we can compute their slopes $\Sigma(\tau)$. (b) $\Sigma(\tau)$ as a function of $ \tau$. In the limit when $\tau$ is larger than the relaxation time of the system $\Sigma(\tau)$ is equal to $M_{\Deltae}/T_r$ (dashed line), as expected from the steady state FT Eq.~(\ref{Symetryfuncdef}). (c) The symmetry function $\ln(P(\Deltau s_{\rm tot})/P(-\Deltau s_{\rm tot}))$ versus the total entropy production $\Deltau s_{\rm tot}$ for different values of $\tau$. As predicted by the detailed FT, the data lie on a straight line with slope one (dashed line). In (a) and (c), the deviations from the linear regime for large fluctuations are due to a lack of statistics in a region where the pdfs are at least two orders of magnitude smaller than their maximum value.}      
\label{work_fluct_noconv}
 \end{center}
\end{figure*}

Based on the Langevin description, the amount of work injected into the rotor during a time $\tau$ is $W_\tau = \int_t^{t+\tau}M_{\Deltae} \omega(t'){\rm d} t' = M_{\Deltae} \Deltau \theta$, where $\Deltau \theta\equiv \theta(t+\tau) - \theta(t)$, i.e., the symbol $\Deltau$ from hereon represents the difference measured over a time delay $\tau$. 
For $W_\tau$ we expect a FT of the form
\begin{equation}
\ln\!\left(\frac{P(W_\tau)}{P(-W_\tau)}\right) = \ln\!\left(\frac{P(\Deltau \theta)}{P(-\Deltau \theta)}\right) = \frac{M_{\Deltae} \Deltau \theta}{T_{\rm r}}\,,\tau\gg\tau_c
\label{Symetryfuncdef}
\end{equation}
where $P(\Deltau \theta)$ is the pdf of $\Deltau \theta$. The expression in the left hand side of the equation is called the symmetry function. Due to the Gaussianity of the distribution of $\omega$, the pdfs of $\Deltau \theta$ are gaussian too, and the symmetry functions are expected to be linear with $\Deltau \theta$ for different values of $\tau$. This is indeed found in experiment (Fig.~\ref{work_fluct_noconv}a). The slope, $\Sigma(\tau)$, is found to decrease with $\tau$ and reaches the constant value $M_{\Deltae}/T_{\rm r}$ expected from the FT Eq.~(\ref{Symetryfuncdef}) for $\tau>3$~s, which is approximatively $3$ times the relaxation time of the system. Thus we find that two independent measurements --one from the pdf of Fig.~\ref{velocity_pdf-psd}b and one using the FT Eq.~(\ref{Symetryfuncdef})-- lead to the very same value of the rotor temperature $T_r$, hereby confirming the validity of the FT Eq.~(\ref{Symetryfuncdef}).

We now turn to the fluctuations of the total (trajectory dependent) entropy production in a time span $\tau$, $\Deltau s_{\rm tot}$, as defined by Seifert in~\cite{Seifert05}. We follow~\cite{Ciliberto2010} and first write the dissipated heat $Q_{\tau}$ \cite{footnoteGRAN} as the injected work $W_\tau$ minus the difference of the kinetic energy of the rotor $E_k=\frac{1}{2} I \omega^2$ at the beginning and end point. The entropy change $\Deltau s_{\rm m}$ in a time span $\tau$ is now defined as the dissipated heat divided by the ratchet temperature
\begin{equation}
T_{\rm r}\, \Deltau s_{\rm m} \equiv Q_{\tau} \equiv W_{\tau} - \Deltau E_k\,.
\label{eq_entropy}
\end{equation}
First we note that for the symmetric system (i.e., without the symmetry-breaking coating) the dissipated heat is equal to $\Deltau E_{k,sym} = T_r \Deltau s_{\rm m, sym}$, which is small but not strictly zero. This contribution has to be subtracted from the entropy change Eq.~(\ref{eq_entropy}) in order to obtain the entropy created by the presence of the external torque $M_{\Deltae}$. As argued in~\cite{Ciliberto2010}, it is this last entropy that may be identified as the total entropy production, i.e., $\Deltau s_{\rm tot} = \Deltau s_{\rm m} - \Deltau s_{\rm m, sym}$.
This quantity is interesting because it should satisfy a detailed FT: The symmetry functions for $\Deltau s_{\rm tot}$ are expected to be equal to $\Deltau s_{\rm tot}$ itself for all values of $\tau$ when the system is in a steady state \cite{Seifert05,Esposito10}. 
From the Langevin equation~(\ref{Langevin_model}), the total entropy takes the following form
\begin{equation}
T_{\rm r}\Deltau s_{\rm{tot}} =M_{\Deltae} \Deltau \theta - I \langle\omega\rangle \Deltau \omega .
\end{equation}
In Fig.~\ref{work_fluct_noconv}c we present the symmetry functions for different integration times $\tau$: We find that the pdfs are Gaussian (not shown) and, most importantly, that the symmetry functions lie on a straight line with slope one within experimental errors, whatever the time delay $\tau$ is. So indeed a detailed FT holds for the total entropy production. 

\emph{State II} -- In state II the dynamics is completely different due to the coupling to a spontaneous convection roll in the granular heat bath: The time series of Fig.~\ref{velocity_pdf-psd}a, shows fluctuations around two preferred velocities, corresponding to the two rolls. As a result, in the pdf of $\omega$ (Fig.~\ref{velocity_pdf-psd}c) two distinct peaks appear. Their different heights reflect that the system obtains a finite mean angular velocity $\langle\omega\rangle$ by residing preferably around the positive peak -- in contrast to the shift of the entire pdf in state I.  
From Fig.~\ref{velocity_pdf-psd}b we find that in both states the psds are almost exactly the same and very well fitted by a Lorentzian. This indicates that the fluctuations in each of the two peaks --and therefore the short time dynamics-- are very similar to those of state I.

This suggests the separation of the dynamics into two parts: On a short time scale, $\omega$ fluctuates around a mean value, with dynamics described by a Langevin equation similar to Eq.~(\ref{Langevin_model}), independent of the direction of the roll. On a much larger time scale, a coupling between the granular gas and the vanes induces collective motion, in which every now and then, through a particularly strong fluctuation, the mean value of $\omega$ quickly switches between positive and negative, reversing both the sense of rotation of the vanes and the roll. These reversals happen randomly in time. The system stays few seconds in each state as can be seen in Fig.~\ref{velocity_pdf-psd}a. Due to the asymmetric coating of the rotor, reversals are easier to realize when the vanes are rotating in the clockwise direction than anti-clockwise, explaining why the system has a preference for the latter~\cite{footnote2}.

Guided by the above considerations, we write the pdf of $\omega$ as the sum of two Gaussian distributions with the same variance $\sigma^2=T_{\rm r}/I$; one with mean value $\omega_0+\omega_{\Deltae}$ and the other with $-\omega_0+\omega_{\Deltae}$:
\begin{eqnarray}
P(\omega)&=& A\left[p_{+}\exp\left(-\frac{I}{2T_{\rm r}}(\omega-(\omega_0+\omega_{\Deltae})^2\right)\right.\nonumber\\&+&\left.p_{-}\exp\left(-\frac{I}{2T_{\rm r}}(\omega-(-\omega_0+\omega_{\Deltae})^2\right)\right]\,.
\label{double_gauss}
\end{eqnarray}
Here $\omega_{\Deltae}$ stands for the shift due to the symmetry breaking coating (i.e., $\pm\omega_0$ denote the locations of the peaks without coating). The factors $p_{\pm}$ represent the weight of each peak (the probability of each direction of the convection roll) and the normalization factor $A = I/(8\pi T_\textrm{r})^{1/2}$ has been chosen such that $p_{+}+p_{-}=1$ implies $\int{P(\omega)d\omega}=1$.

\begin{figure}
\centering
\includegraphics[width=0.8\linewidth]{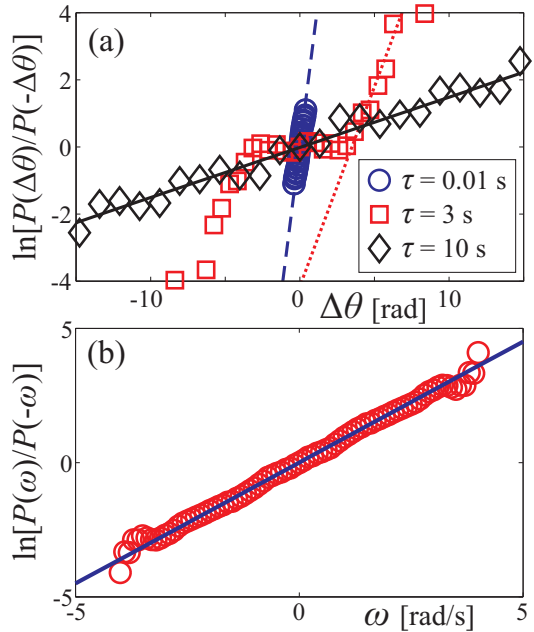}
\caption{\small (color online) State II [$S=2.15$ and $h = 51$~mm]: (a) Symmetry function of $\Delta \theta$ for three different integration times. (b) Symmetry function of $\omega$. } \label{FT_convection}
\end{figure}

What can we learn from FTs for this very non-Gaussian system? It seems reasonable to assume that if a FT exists, it will be for the work done on short timescales, when the system resides in either of the two peaks of the pdf of $\omega$. And indeed, in Fig.~\ref{FT_convection}a (dashed blue line) we find that the symmetry function for $\Deltau\theta$ is linear in $\Deltau\theta$ for $\tau = 0.01$ s. Since in this limit $\omega \approx \Deltau\theta/\tau$, we may expect a similar symmetry relation for $\omega$, which we plot in Fig.~\ref{FT_convection}b. Indeed, $\ln{[P(\omega)/P(-\omega)]} \propto \omega$, which is truly remarkable in view of the peculiar shape of its distribution (Fig.~\ref{velocity_pdf-psd}c). Computing the symmetry function from the double Gaussian of Eq.~(\ref{double_gauss}) we find that it can only be linear if the weights of the peaks obey  
\begin{equation}
\frac{p_+}{p_-} = \exp\left(\frac{2I\,\omega_{\Deltae}\,\omega_0}{T_{\rm r}}\right)\,,
\label{eq_ratio}
\end{equation}
which then directly leads to the symmetry relation 
\begin{equation}
\ln\frac{P(\omega)}{P(-\omega)} = \frac{2I\omega_{\Deltae}}{T_r} \omega \,.
\end{equation}
From the slope of the experimental symmetry function (Fig.~\ref{FT_convection}b), which, with $T_r = 0.062$ mJ leads to $\omega_{\Deltae} = 0.372 rad/s$ rad/s, combined with $\omega_0 = 1.8$ rad/s from the pdf of the symmetric system, we compute $p_+$ and $p_-$ and plot the resulting theoretical pdf in Fig.~\ref{velocity_pdf-psd}c. The agreement with the experimental pdf is good, except for angular velocities close to zero.   

We now return to the symmetry function of $\Deltau\theta$ in Fig.~\ref{FT_convection}a): For the smallest value ($\tau=0.01$ s) we find a symmetry relation. This is because the changes $\Deltau\theta$ are dominated by the fluctuations, and the reversals are just rare events without large consequences for the value of $\Deltau\theta$. For the intermediate value $\tau =  3$ s, there is no valid symmetry relation. This stands to reason because when the integration time $\tau$ increases, the probability that it contains at least one reversal increases as well. $\Deltau\theta$ is then likely to include a reversal and there is no reason to consider the system to be in a steady-state, which is a prerequisite for a symmetry relation to hold. There are two distinct regimes: One at small $| \Deltau\theta |$, dominated by reversals, and another at large $| \Deltau\theta | \gtrsim \omega_0\tau $, a distance which in the given time interval is unlikely to be reached when the trajectory includes one or more reversals (dotted red line in Fig.~\ref{FT_convection}a). 
For very large $\tau$, typically many reversals are included in each $\Deltau\theta$ and we can consider the system to be in a steady state again. Indeed, for $\tau = 10$ s the pdf is Gaussian (not shown) due to the central limit theorem, and the symmetry function tends to become linear (Fig.~\ref{FT_convection}a, solid black line). However, the experimental slope is smaller than the slope $M_{\Deltae}/T_r$ one would expected based upon Eq.~(\ref{Symetryfuncdef}). This may be because the entropy production depends on two parts: one is the work $M_{\Deltae}\Deltau\theta$ and a second one coming from another variable responsible for the reversals, which is not included here (cf. the theoretical work in~\cite{Lacoste09}).

In conclusion, we have investigated the validity of FTs in two different states of an asymmetric rotor experiment in a granular gas. The first state, with a Gaussian pdf for the angular velocity, is found to be well described by a first order Langevin equation and therefore analogous to a Brownian system of temperature $T_{\rm r}$. This is reflected in the observation that the FT is valid at large $\tau$ for the injected work and for all $\tau$ for the total entropy production. In the second state the pdf of $\omega$ is double-peaked due to convection rolls developing in the granular gas. Here symmetry relations are found for very large and very small $\tau$, of which the latter of leads to a good prediction of the ratio of the height of the two peaks. 

The authors thank S. Ciliberto for helpful discussions. The work is part of the research program of FOM, which is financially supported by NWO.


\begin{thebibliography}{0}
\bibitem{footnote1}
Theoretically, the ratchet temperature is related to the granular temperature by $T_{\rm r} = T_{\rm g}(1+e)/2$. Both are orders of magnitude larger than $k_B T$.
\bibitem{footnoteGRAN} It should be stressed that here heat and entropy are not the thermal quantities but are defined by analogy to those for the stochastic motion of the rotor, i.e., with the temperature $T_r$.

analogous with the corresponding
\bibitem{footnote2}
Without the asymmetric coating, both directions are (within experimental and statistical errors) found to have the same probability.
\end{thebibliography}
\end{document}